\documentclass[twocolumn]{aastex63}

\usepackage{amsmath}
\usepackage{graphicx}
\usepackage{natbib}
\usepackage[caption=false]{subfig}
\usepackage[nameinlink]{cleveref}
\usepackage{verbatim}
\usepackage[caption=false]{subfig}

\newcommand\gaia{{\it Gaia}}
\newcommand{\pmra}{$\mu_{\alpha{*}}$}
\newcommand{\pmdec}{$\mu_\delta$}

\received{}
\revised{}
\accepted{}
\submitjournal{ApJ}

\shorttitle{LMC Southern Periphery Substructure}
\shortauthors{Cheng et al.}
\graphicspath{{./}{figures/}}
\begin{document}

\title{Kinematical Analysis of Substructure in the Southern Periphery of the Large Magellanic Cloud}

\author[0000-0002-7009-3957]{Xinlun Cheng}
\affiliation{Department of Astronomy, University of Virginia, Charlottesville, VA 22904-4325, USA}

\author[0000-0003-1680-1884]{Yumi Choi}
\affiliation{Space Telescope Science Institute, 3700 San Martin Drive, Baltimore, MD 21218, USA}

\author[0000-0002-7134-8296]{Knut Olsen}
\affiliation{National Optical-Infrared Astronomy Research Laboratory (NOIRLab), 950 North Cherry Avenue, Tucson, AZ 85719, USA}

\author[0000-0002-1793-3689]{David L. Nidever}
\affiliation{Department of Physics, Montana State University, P.O. Box 173840, Bozeman, MT 59717, USA}

\author[0000-0003-2025-3147]{Steven R. Majewski}
\affiliation{Department of Astronomy, University of Virginia, Charlottesville, VA 22904-4325, USA}

\author[0000-0003-2325-9616]{Antonela Monachesi}
\affiliation{Instituto de Investigaci\'on Multidisciplinar en Ciencia y Tecnolog\'ia, Universidad de La Serena, Ra\'ul Bitr\'an 1305, La Serena, Chile}
\affiliation{Departamento de Astronom\'ia, Universidad de La Serena, Av. Cisternas 1200, La Serena, Chile}

\author[0000-0003-0715-2173]{Gurtina Besla}
\affiliation{Department of Astronomy, University of Arizona, 933 North Cherry Avenue, Tucson, AZ 85721, USA}

\author[0000-0002-8129-3487]{C\'esar Mu\~noz}
\affiliation{Instituto de Investigaci\'on Multidisciplinar en Ciencia y Tecnolog\'ia, Universidad de La Serena, Ra\'ul Bitr\'an 1305, La Serena, Chile}
\affiliation{Departamento de Astronom\'ia, Universidad de La Serena, Av. Cisternas 1200, La Serena, Chile}

\author[0000-0001-5261-4336]{Borja Anguiano}
\affiliation{Department of Astronomy, University of Virginia, Charlottesville, VA 22904-4325, USA}

\author{Andres Almeida}
\affiliation{Department of Astronomy, University of Virginia, Charlottesville, VA 22904-4325, USA}

\author{Ricardo R. Mu\~noz}
\affiliation{Departamento de Astronom{\'i}a, Universidad de Chile, Camino El Observatorio 1515, Las Condes, Chile}

\author{Richard R. Lane}
\affiliation{Centro de Investigaci{\'o}n en Astronomía, Universidad Bernardo O'Higgins, Avenida Viel 1497, Santiago, Chile}

\author[0000-0003-4752-4365]{Christian Nitschelm}
\affiliation{Centro de Astronom{\'i}a (CITEVA), Universidad de Antofagasta, Avenida Angamos 601, Antofagasta 1270300, Chile}

\begin{abstract}
We report the first 3-D kinematical measurements of 88 stars in the direction of several recently discovered substructures in the southern periphery of the Large Magellanic Cloud (LMC) using a combination of {\it Gaia} proper motions and radial velocities from the APOGEE-2 survey. More specifically, we explore stars lie in assorted APOGEE-2 pointings in a region of the LMC periphery where various overdensities of stars have previously been identified in maps of stars from {\it Gaia} and DECam. By using a model of the LMC disk rotation, we find that a sizeable fraction of the APOGEE-2 stars have extreme space velocities that are distinct from, and not a simple extension of, the LMC disk. Using N-body hydrodynamical simulations of the past dynamical evolution and interaction of the LMC and Small Magellanic Cloud (SMC), we explore whether the extreme velocity stars may be accounted for as tidal debris created in the course of that interaction. We conclude that the combination of LMC and SMC debris produced from their interaction is a promising explanation, although we cannot rule out other possible origins, and that these new data should be used to constrain future simulations of the LMC-SMC interaction. We also conclude that many of the stars in the southern periphery of the LMC lie out of the LMC plane by several kpc. Given that the metallicity of these stars suggest they are likely of Magellanic origin, our results suggest that a wider exploration of the past interaction history of the Magellanic Clouds is needed.

\end{abstract}


\section{Introduction} \label{sec:intro}
As the closest interacting pair of dwarf galaxies, the Large and Small Magellanic Clouds (LMC and SMC) are excellent laboratories for exploring dwarf galaxies and their interaction in detail. Consequently, the Clouds have been the targets of many dedicated observational campaigns. In particular, recent large and contiguous imaging surveys have accelerated discoveries of low surface brightness stellar substructures around the Magellanic periphery \citep[e.g.,][]{Mackey2016, Pieres2017, Mackey2018, Belokurov2019, Martinez-Delgado2019, Gaia2021}, made possible by virtue of, for example, Gaia \citep{Gaia2016}, the DECam/Blanco surveys \citep[e.g.,][]{DES2005, Nidever2017,DELVE}, and work with the VISTA facility \citep[e.g.,][]{Cioni2011, ElYoussoufi2021}. 

These outlying stellar substructures in the Magellanic periphery are sensitive probes for deciphering the tidal interaction histories between the LMC and SMC and between the Clouds and the Milky Way (MW) because the shallower potentials in galactic peripheries make stars there more easily disturbed. Thus, identifying low surface brightness stellar substructures in the LMC and SMC outskirts and measuring their key properties is essential for understanding their dynamics. While some studies of the morphology and stellar populations of these faint structures have been conducted \citep[e.g.,][]{Mackey2018, Martinez-Delgado2019, ElYoussoufi2021}, the detailed 3D kinematics for those structures remain largely unexplored.

One of the prominent stellar substructures around the LMC is an arm-like feature in the northern periphery \citep{Mackey2016}. \citet{Cullinane2021} showed that the stellar metallicity and kinematics of this northern arm are consistent with those of the outer LMC disk and attributed the formation of the northern arm to the MW tide. Given that many of the stellar structures in the main body of the LMC are found to be asymmetric --- for example, a one-armed spiral and an off-centered bar \citep{deVaucouleurs1972}, as well as two stellar warps seen only in the southwest part of the disk \citep{Olsen2002, Choi2018a} --- it is important to determine whether the northern arm is yet another asymmetric feature of the LMC or if it has a still-unidentified counterpart in the southern periphery. If a counterpart indeed exists, it would place constraints on formation mechanisms for these particular features, which, in turn, are a key to deciphering the LMC's interaction histories with the SMC and MW.

Recently, a candidate counterpart of the northern arm was discovered in the southern periphery by \citet{Belokurov2019}. Based on N-body simulations of the Magellanic Clouds that included the MW potential, they suggested that the southern structure is likely a spiral arm created by the most recent interaction with the SMC and consisting of pulled-out LMC disk stars. These authors further suggest that the stellar motions in the southern structure retain the kinematic signature of the outer LMC disk. Their assessment, however, was based on 2D proper motion measurements, not the full 3D velocity information that is essential to making confident conclusions regarding the  origin of these stars.

Another prominent stellar substructure in the southern part of the LMC periphery are two large ``hook''-like features, discovered by \citet{Mackey2018} and designated as ``Substructure 1'' and Substructure 2'' in their paper. These ``hook''-like features reside to the south of the LMC's main disk at $\sim$10$^\circ$ from the LMC center, with $\sim$40--45$^\circ$ separation in position angle between them (see \Cref{fig:lmc_dist}). Based on a comparison of the relative color-magnitude diagram (CMD) positions of the red clump and main-sequence turnoff stars in these regions, \citet{Mackey2018} concluded that the distances to the ``hook''-like features are not significantly different from those of the stars in both the northern and southern peripheries. Mackey et al. also suggest a physical association between ``Substructure 2'' and the RR Lyrae Bridge \citep{Belokurov2017} connecting the LMC and SMC. However, no kinematical studies have been conducted on these substructures to date. 

In this study, we explore the kinematics of stellar substructures around the LMC, with particular focus on the southern periphery, including  ``Substructure 1'' and ``Substructure 2'' (i.e., two ``hook''-like features). We make use of the improved uncertainties in proper motion measurements from \gaia \ Early Data Release 3 \citep[EDR3;][]{Gaia2016, gaiaedr3} and new radial velocity measurements from APOGEE spectra.  We are guided in our interpretation of these features by a model of LMC rotation that we have developed, as well as various N-body simulations \citet{Besla2012} of the dynamical history and past interaction of the LMC and SMC, which produce a variety of perturbations and tidal debris from either or both of the Clouds, depending on starting assumptions. 
The APOGEE spectra also allow us to investigate the added dimension of the stellar metallicity distributions of these substructures, further clues to their origin.  A companion exploration (Mu\~noz et al., in preparation) with these same spectroscopic data will focus on the detailed chemical aspects of these substructures to further constrain the properties and origin of the stellar substructures in the southern periphery.

This paper is organized as follows: In Section~\ref{sec:data}, we describe the MC star samples used in this study. In Section~\ref{sec:results}, we present the 2D and 3D stellar motions for stars using only Gaia proper motions (PMs) and Gaia PMs plus APOGEE radial velocities, respectively. We particularly focus on a kinematically distinct group of stars that lie around, but are not limited to, the southern structures discovered by \citet{Belokurov2019}. We then present comparisons with hydrodynamical simulations of an LMC-SMC analog pair of galaxies to explore plausible explanations for those kinematically distinct stars in the southern periphery.  In Section~\ref{sec:conclusion}, we discuss the possible origin of these newfound MC stellar substructures and summarize our conclusions. 

\section{Data} \label{sec:data}
\begin{figure*}
    \centering
    \subfloat[]{
        \includegraphics[width=0.49\textwidth]{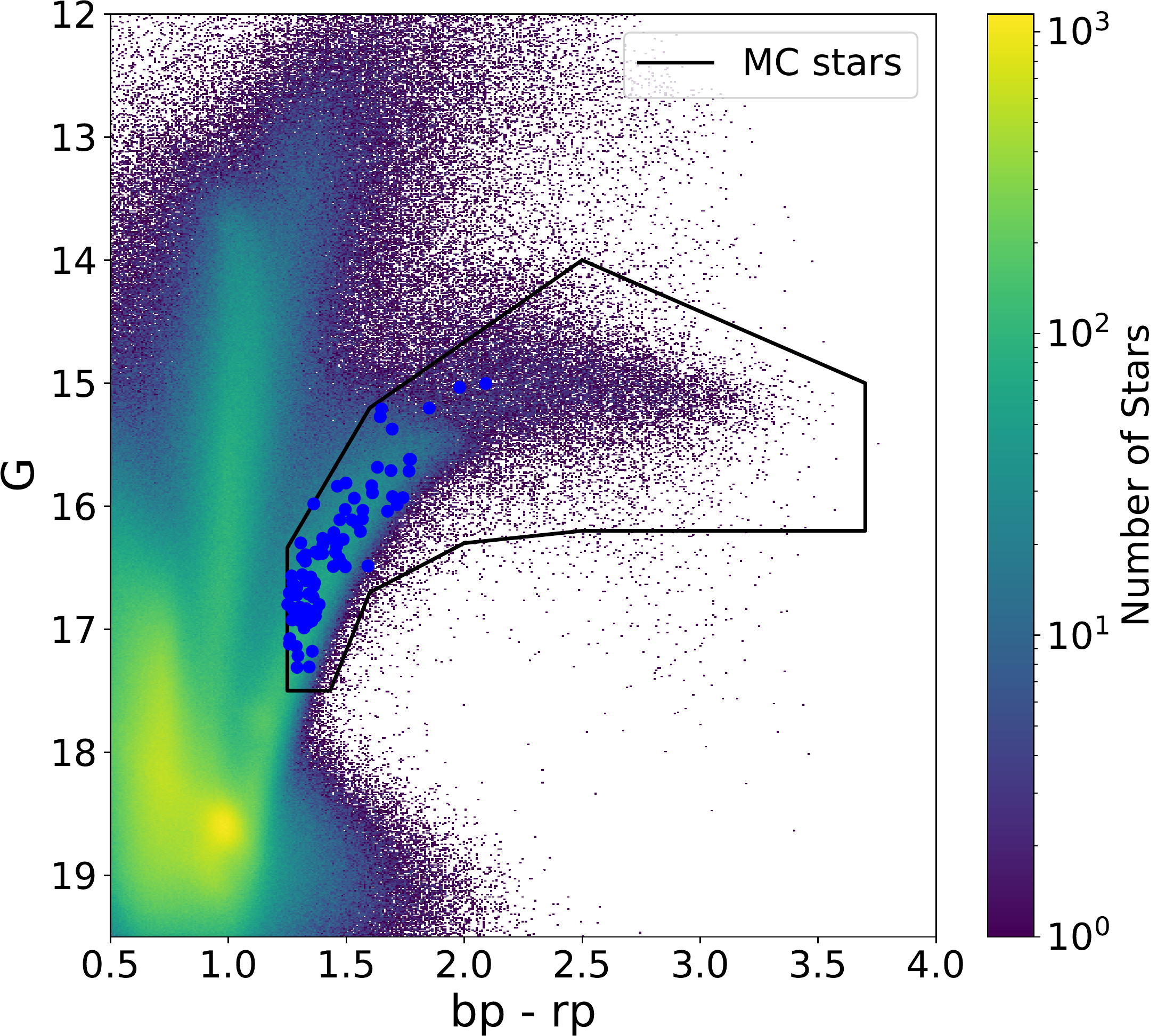}
        \label{fig:cmd_selection}
    }
    \hspace*{\fill}
    \subfloat[]{
        \includegraphics[width=0.49\textwidth]{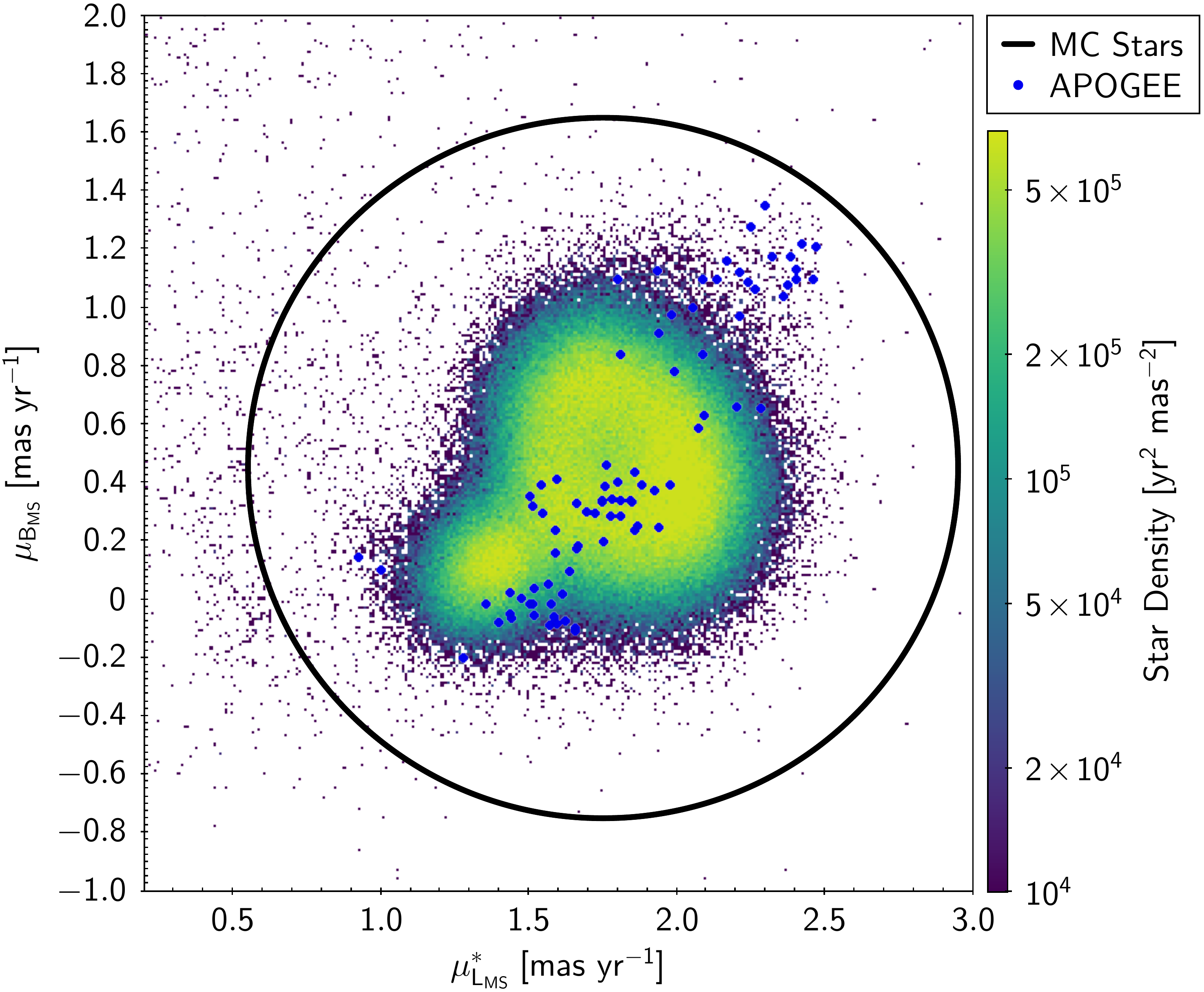}
        \label{fig:pm_selection}
    }\\[2ex]
    \subfloat[]{
        \includegraphics[width=0.7\textwidth]{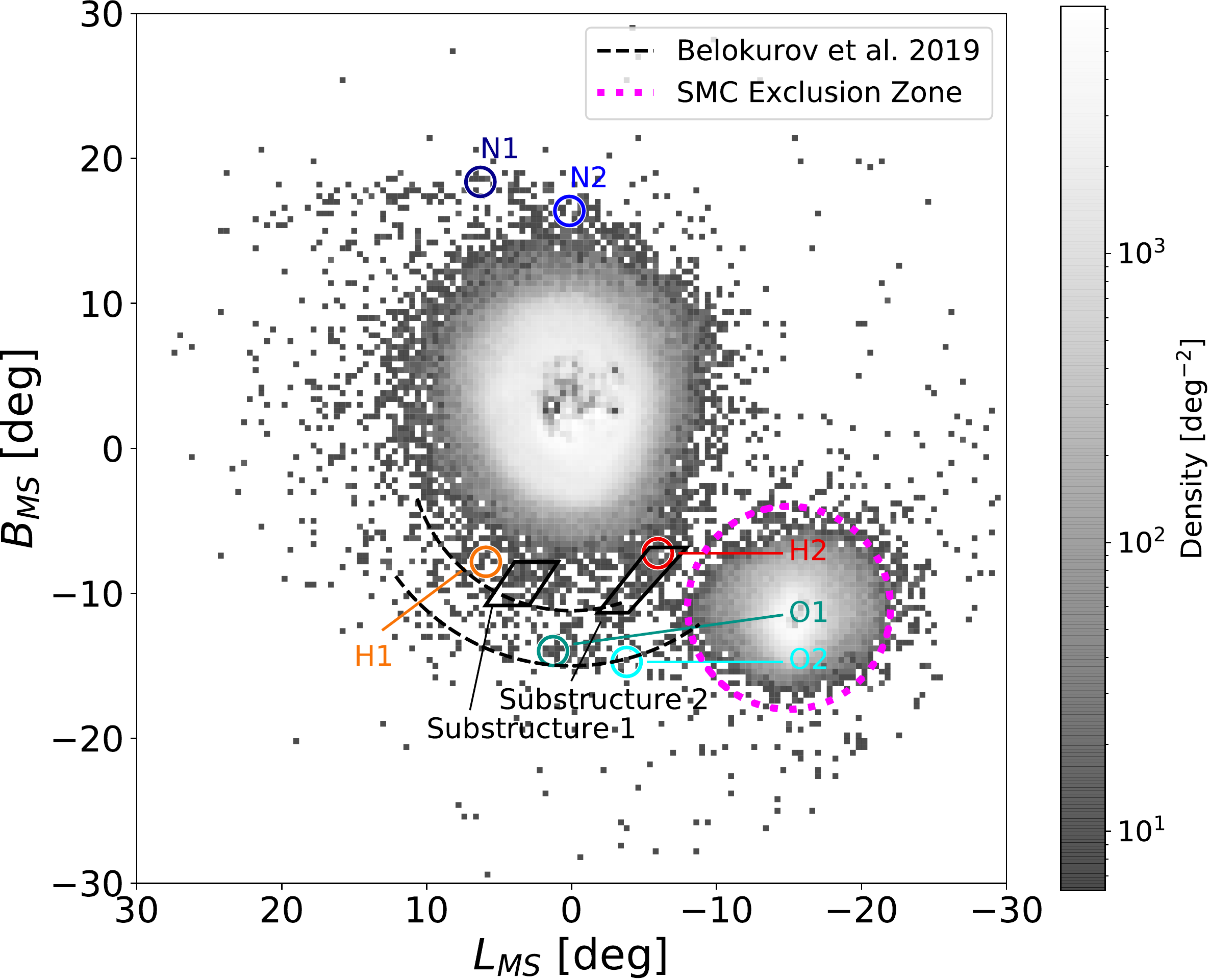}
        \label{fig:lmc_dist}
    }
    \caption{The steps to creating our survey sample.  (a) The CMD of all {\it Gaia} stars within 30$^{\circ}$ of the origin of the Magellanic Stream coordinate system. The black lines indicate the CMD region showing our initial selection of LMC stars. APOGEE stars are indicated with blue dots. (b) Proper motions in Magellanic Stream coordinates and our proper motion selection criteria. The black lines indicate the regions within which we retain stars in our sample.  (c) On-sky distribution of our selected LMC star sample in Magellanic Stream coordinates ($L_{MS}$, $B_{MS}$). The locations of APOGEE fields are indicated with circles of different colors. Substructures identified in \citet{Mackey2018} are identified and labelled within solid black lines, and substructures (arms) identified in \citet{Belokurov2019} are labelled with dotted black lines.}\label{fig:selection}
\end{figure*}

Our analysis relies on data from \gaia \ EDR3 \citep{gaiaedr3}, from which we draw LMC stars via a selection procedure similar to that applied by \citet{Belokurov2019}, but with some slightly different criteria: Stars with $G<17.5$ are selected within 30$^\circ$ of the origin of the Magellanic Stream ($\alpha=80.8926^\circ$, $\delta=-72.1859^\circ$) coordinate system \citep{Nidever2008}. We adopted the extinction map from \citet{Schlafly2011}, and an extinction correction is performed with the equation and parameters from \citet{GaiaHR}.  Then we make a selection within the color-magnitude diagram to constrain our sample to stars primarily along the red giant branch (RGB) of the LMC (\Cref{fig:cmd_selection}). To eliminate most of the foreground stars from the Milky Way, those with parallax $\varpi>0.2\ \text{mas}$ or Galactic latitude $|b|<5^\circ$ are removed, while an additional selection for stars with similar proper motion to the LMC is applied (\Cref{fig:pm_selection}; in this figure, proper motions are shown in Magellanic Stream coordinate system, and the large and small ``blobs'' represent stars from the LMC and SMC, respectively.) The spatial distribution of our selected LMC star sample is shown in \Cref{fig:lmc_dist}. In \citet{Mackey2018}, two substructures to the south of LMC have been identified as regions of stellar overdensity; for ease of comparison, these substructures are labeled in \Cref{fig:lmc_dist} and subsequent figures as Substructure 1 and Substructure 2. Similarly, structures identified by \citet{Belokurov2019} are labeled with black dotted line in all relevant figures. Furthermore, we excluded stars within 7 degrees from the center of SMC (red dotted line, SMC Exclusion Zone) from all analysis, but we decide to include these stars within our figures for easy comparison.

To investigate the kinematics of substructure at the southern periphery of the LMC further, we employed stars from APOGEE \citep{Majewksi2017, Wilson2019, GarciaPerez2016, Holtzman2015, Nidever2015, Zasowski2017} Data Release 17 (DR17), part of the Sloan Digital Sky Survey (SDSS-IV) \citep{Gunn2006, Blanton2017}, where precise line-of-sight velocities enable the derivation of the three-dimensional motions of stars. We focus here on six APOGEE fields placed on and around previously known substructures: two to the North of LMC on the arm feature discovered by \citet{Belokurov2019} and four to the South of LMC on the hook features and their extensions discovered by \citet{Mackey2018}; data on these fields were obtained through the Chilean  National  Telescope  Allocation  Committee (CNTAC) program  CN2019A-30 (PI: A. Monachesi). These fields are shown in \Cref{fig:lmc_dist} with circles of different colors that, in some following figures, will be used to identify the stars in each field. We applied the same parallax, color-magnitude diagram, and proper motion selections with our \gaia\, sample to the stars within the six APOGEE fields. Additional selection criteria in line-of-sight velocity ($100<V_{helio}<350$ km s$^{-1}$), effective temperature ($T_{eff}<5400$ K) and surface gravity ($\log{g}<4$) are applied to refine our LMC sample further. A total 88 stars across all 6 fields passed through all selection criteria.

\section{Results} \label{sec:results}
\subsection{2D motion from Gaia EDR3}\label{sec:2Dmotion}

\begin{figure*}
    \centering
    \subfloat[]{
        \includegraphics[width=0.49\textwidth]{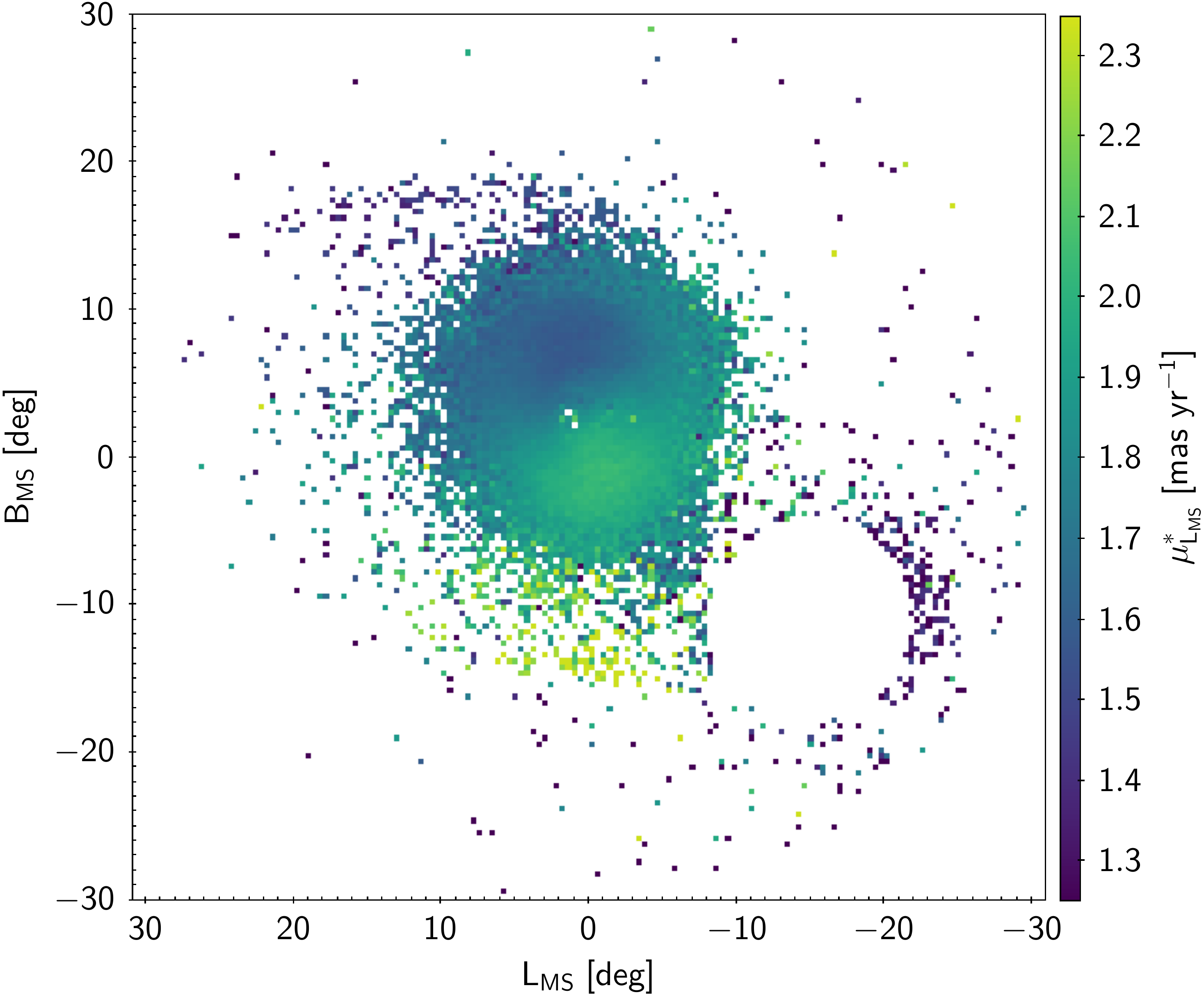}
        \label{fig:lmc_pml}
    }
    \hspace*{\fill}
    \subfloat[]{
        \includegraphics[width=0.49\textwidth]{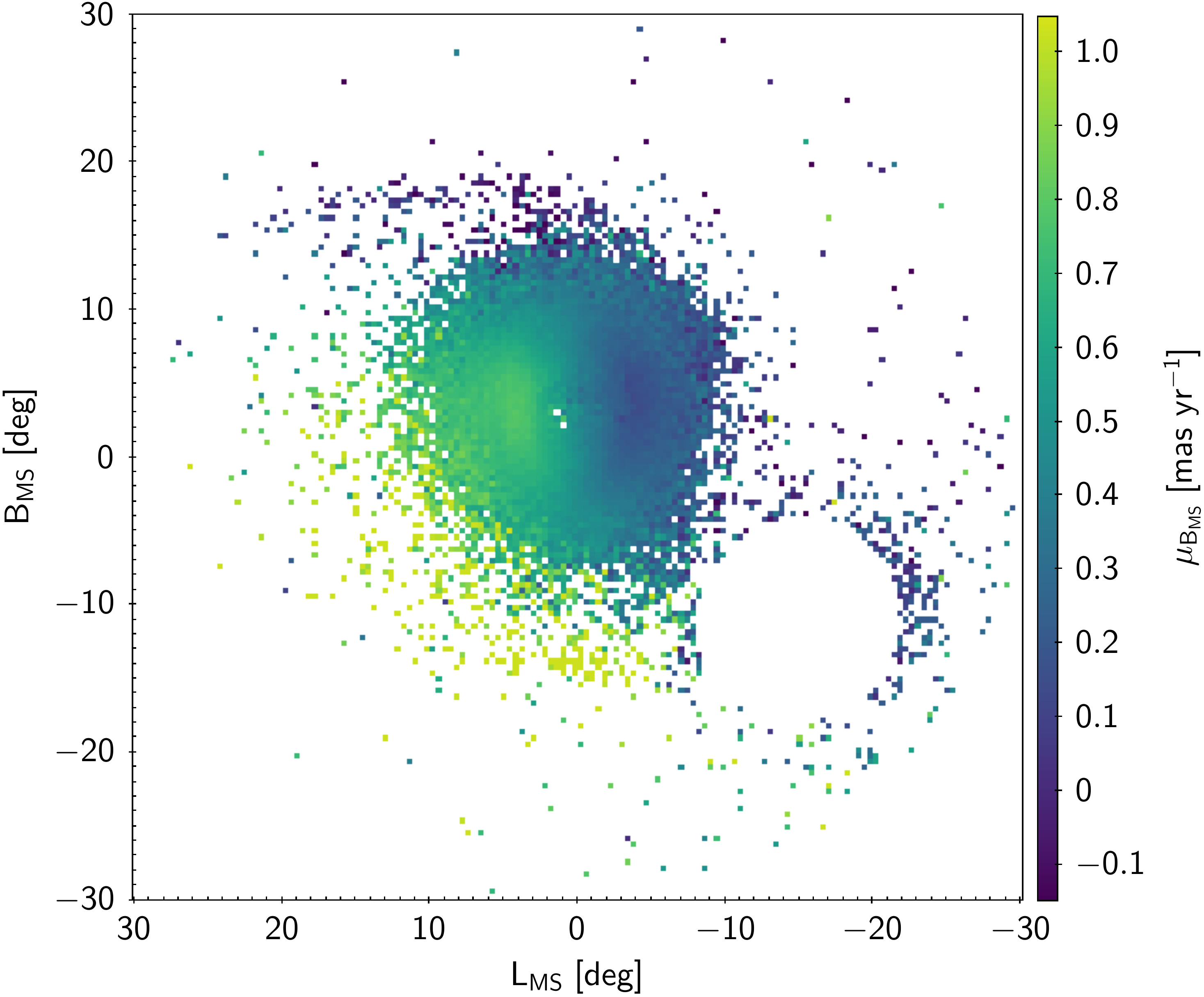}
        \label{fig:lmc_pmb}
    }\\[-2ex]
    \subfloat[]{
        \includegraphics[width=0.49\textwidth]{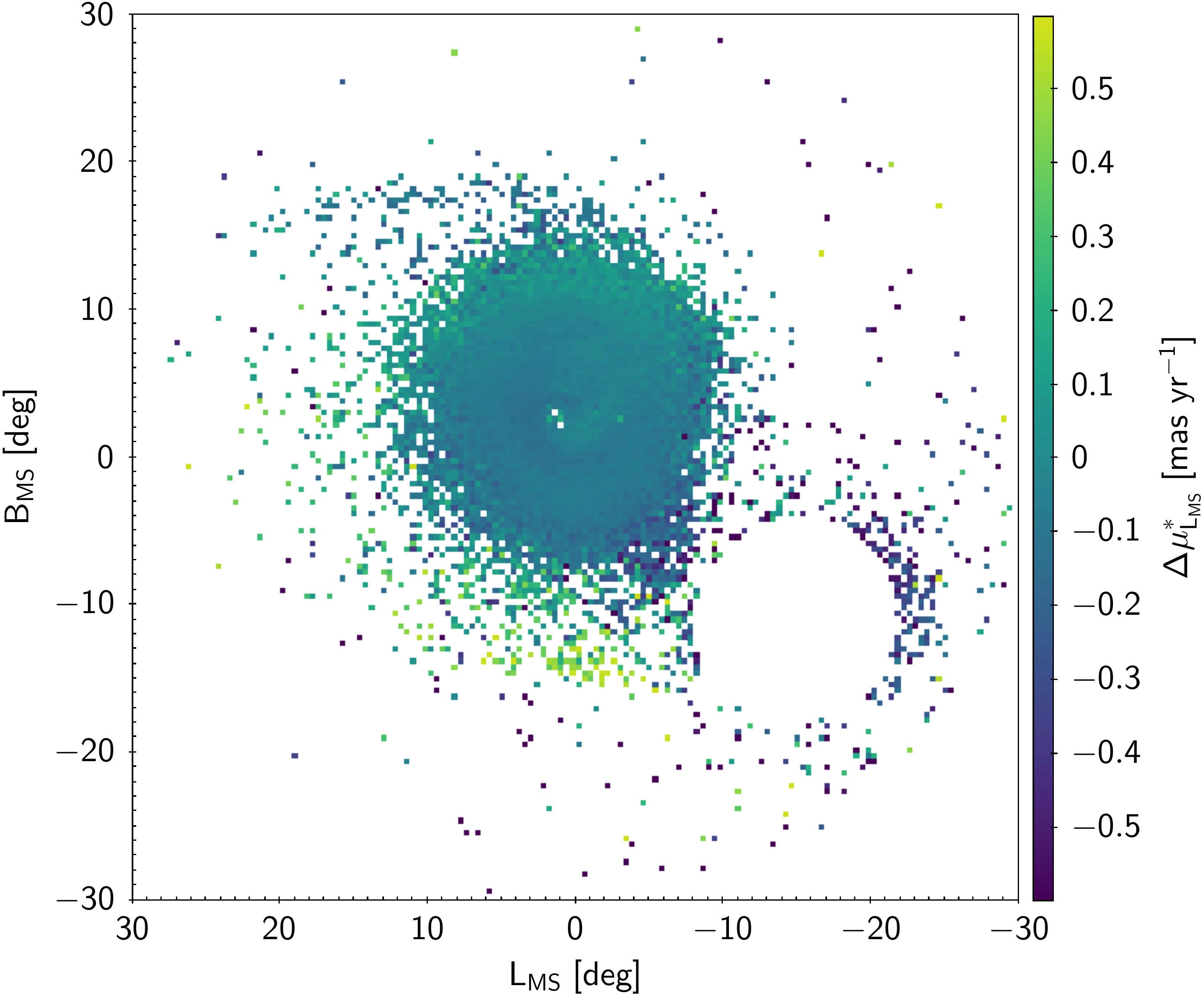}
        \label{fig:lmc_delta_pml}
    }
    \hspace*{\fill}
    \subfloat[]{
        \includegraphics[width=0.49\textwidth]{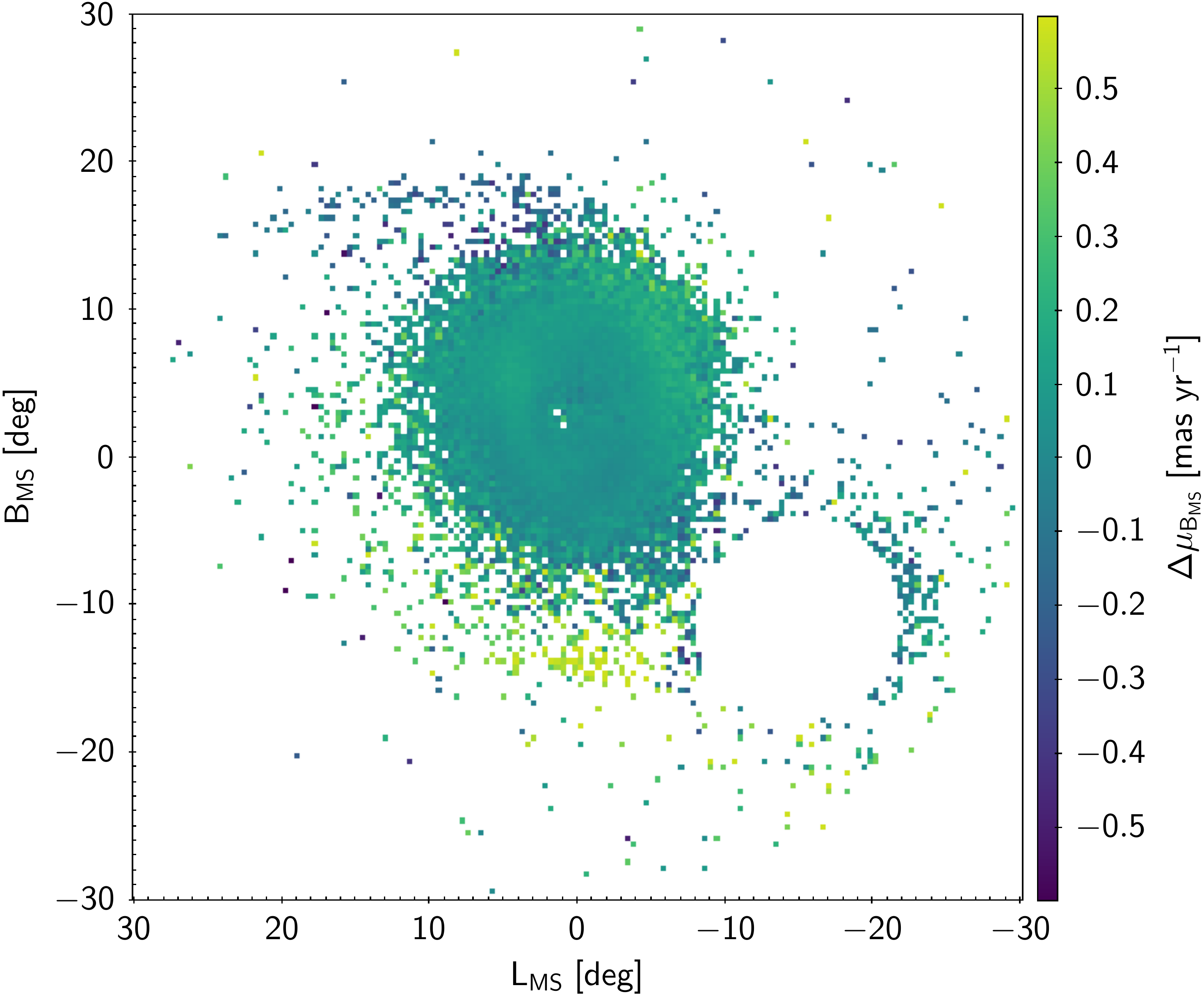}
        \label{fig:lmc_delta_pmb}
    }
    \caption{The observed proper motion distribution of the LMC star sample in Magellanic Stream coordinates (panels a and b). The reported longitudinal proper motion, $\mu^*_{L_{MS}}$, has the $\cos{(B)}$ correction applied. In panels (c) and (d) the observed proper motions are differenced against those predicted in the LMC kinematic model ($\Delta\mu=\mu_{obs}-\mu_{model}$, see Section \ref{sec:2Dmotion}).}\label{fig:lmc_pm}
\end{figure*}

In \Cref{fig:lmc_pm} we show the spatially averaged proper motion components in Magellanic Stream coordinates of the {\it Gaia} sample as observed (top panels), as well as the residuals of those proper motions after subtracting the predicted values from a fitted model (bottom panels), which we describe further below.  The dipole pattern seen in the main LMC disk in the upper panels steems from the disk's rotation, while distinct kinematical signatures of previously discovered features, such as the arm-like substructure to the north of the LMC \citep{Mackey2016} and the hook-like feature lying in between the LMC and SMC (``Substructure 2'' in \citealt{Mackey2018}), are revealed around the LMC periphery. The residual maps in the lower panels show that our kinematical model fits the disk rotation well, whereas the features on the periphery exhibit strong departures in proper motion from our simple model of a rotating disk. In particular, as previously found by Mackey et al., while the substructure to the north features low proper motions, stars in the hook in the south have a much larger proper motion than stars in the immediately surrounding area. In addition, in the southern part of the LMC, at radii extending beyond about 10 degrees from LMC center and starting near the end of the hook and wrapping clockwise around the LMC to about ($L_{MS}$, $B_{MS}$) = ($10^\circ$, $-5^\circ$), there is a swath of stars  that shows higher proper motions in both the longitudinal and latitudinal dimensions. 

To explore these kinematical structures of the LMC periphery further, we contrast the observed motions against those from a kinematical model of the LMC that includes the effects of bulk center-of-mass motion and internal rotation, as described in \citep[submitted]{Choi2022}.
This model is based on fits to $\sim$10$^4$ LMC disk stars with both proper motion measurements from \gaia{} EDR3 \citep{gaiaedr3} and line-of-sight velocity measurements from a variety of sources, including Hydra-CTIO observations of 4226 stars by \citet{Olsen2011}, 556 unpublished Hydra-CTIO observations (Olsen et al.\ in prep), and 5386 stars from SDSS DR16/APOGEE-2 \citep{SDSSDR16}. 
In brief, the modeling procedure, which is based on the formalism of \citet{vanderMarel2002}, fits several parameters jointly to the proper motion and line-of-sight velocity data. These parameters include the location of the LMC's kinematical center in RA and Dec, the LMC's bulk transverse motion along the RA and Dec axes, the line-of-sight velocity of the kinematical center, the position angle of the line of nodes, the inclination of the disk, two parameters describing the shape and amplitude of an internal rotation curve that is flat after a scale radius $R_0$, and the velocity dispersion in three orthogonal directions. We assume that the LMC disk has no precession or nutation and that the distance to the LMC is 50.1 kpc \citep{Freedman2001}. The model predicts the proper motion distribution well within the inner disk of the LMC, but does significantly deviate from the observations at larger radii \citep[submitted]{Choi2022}, and (as expected) fits especially poorly to stars in the periphery of the SMC.

\begin{figure*}
    \centering
    \subfloat[]{
        \includegraphics[width=1.2\columnwidth]{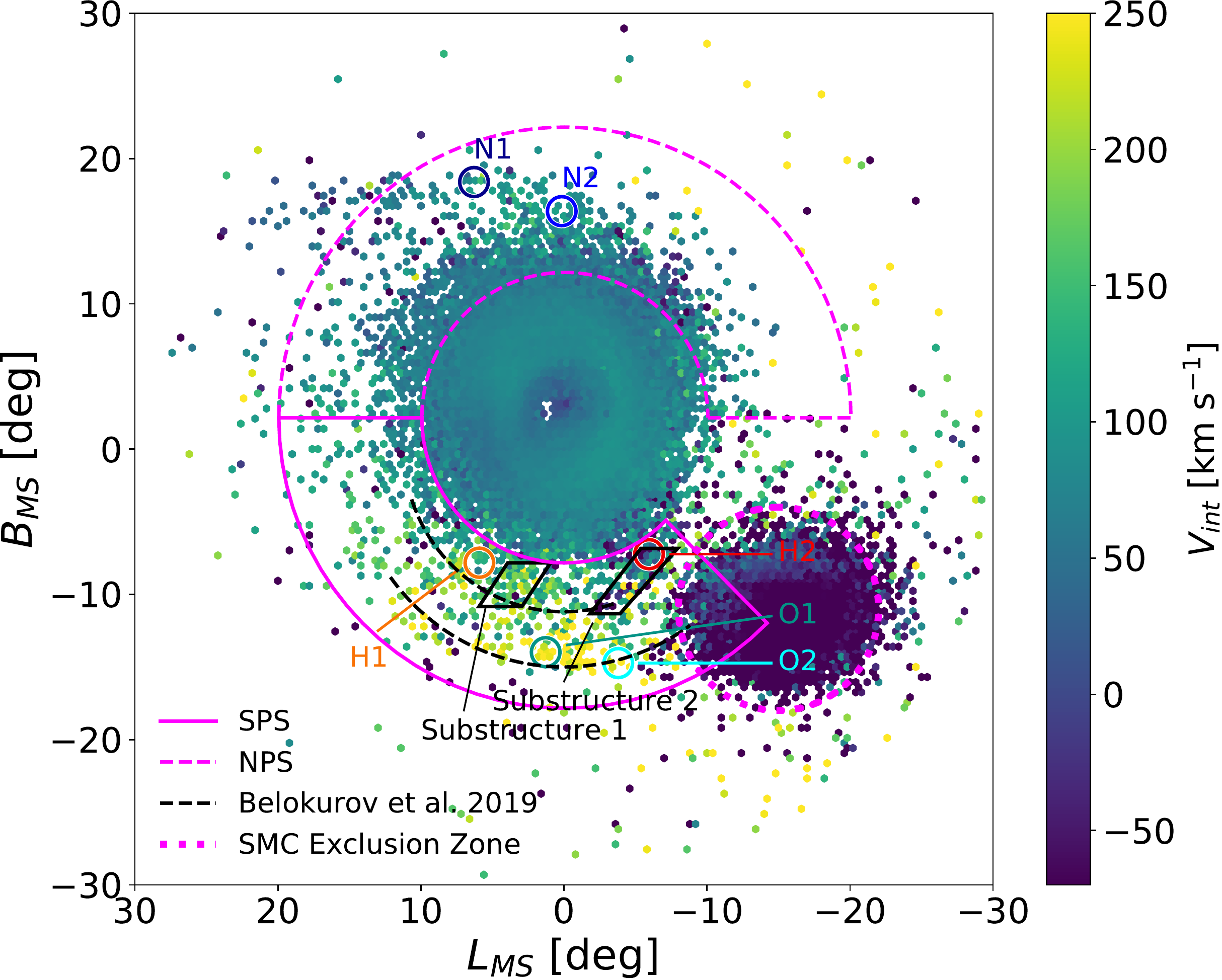}
        \label{fig:lmc_vint}
    }
    \hspace*{\fill}
    \subfloat[]{
        \includegraphics[width=0.8\columnwidth]{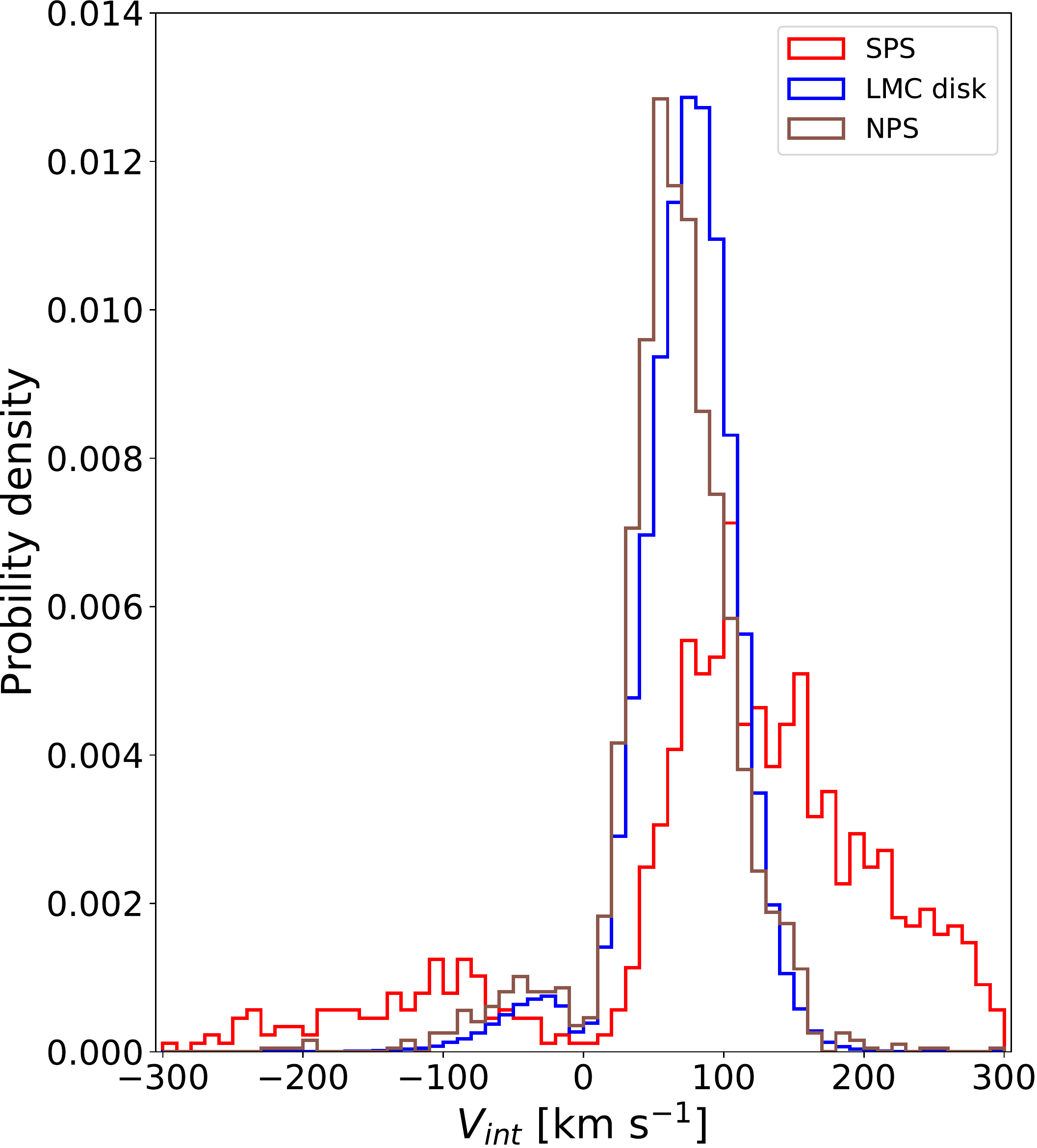}
        \label{fig:lmc_vint_dist}
    }
    \caption{In-plane velocity distributions for our selected LMC star sample. (a) The in-plane velocity distribution, $V_{int}$ in Magellanic Stream coordinates ($L_{MS}$, $B_{MS}$). Some previously identified substructures are indicated, as is the placement of the APOGEE-2 fields and the southern periphery sector (SPS) analyzed separately. A northern periphery sector (NPS) is placed to the north of LMC, with the same inner and outer radius as SPS. (b) A comparison of the $V_{int}$, in-plane velocity distributions for stars in the nominal disk of the LMC (radii less than $8^{\circ}$ from LMC center, blue curve), stars in NPS region (brown curve) to those in the SPS region. 
    }
\end{figure*}

To probe the possible origins of the previously discussed features more deeply, we use the model fit as described above to deproject the proper motions into in-plane velocity, $V_{int}$, as shown in \Cref{fig:lmc_vint}.  To derive an expression for $V_{int}$, we use the coordinate system and formalism developed by \citet{vanderMarel2002}, in particular: (1) their Equation 7, which describes the relationship between the proper motion vector and the orthogonal velocity components $v_2$ and $v_3$ in the plane of the sky (as defined in Equation 1 of \citet{vanderMarel2002}, the direction of $v_2$ is parallel to radius vector originating at the LMC center and ending at the sky coordinate in question, while $v_3$ is orthogonal to this in the direction of the position angle $\Phi$), and (2) their Equation 21, which describes the projection of the rotation curve to $v_2$ and $v_3$. The distances to the individual stars assume that they are moving in the inclined plane of the LMC disk, and as such depend on the distance to the center of mass of the LMC and on the disk inclination.  We adopted 18.5 as the LMC distance modulus (which is within 1\% of the measurement by \citealt{Pietrzynski2019} from eclipsing binaries) and derived the inclination from the model fit, which we found to be 22.7 $\deg$, in close agreement with that derived from red clump distances by \citet{Choi2018a}.
We use the observed proper motion vector, after subtracting the contribution from center of mass motion, to compute $v_{2, int}$ and $v_{3, int}$, and then derive an expression for the rotation velocity $V_{int}$ as a function of the magnitude of the velocity vector in the plane of the sky:

\begin{equation}
    V_{int} = s\frac{(v_{2, int}^2 + v_{3, int}^2)^\frac{1}{2}}{[(f_1/f_2)^2\sin^2(i)\cos^2(\Phi-\Theta)+f_2^2]^\frac{1}{2}}
\end{equation}
where
\[
    f_1 = \cos(i)\sin(\rho)+\sin(i)\cos(\rho)\sin(\Phi-\Theta)
\]
and 
\[
    f_2 = [\cos^2(i)\cos^2(\Phi-\Theta)+\sin^2(\Phi-\Theta)]^\frac{1}{2}
\]
are terms in the geometric projection and $s=\pm1$ is the direction of orbital motion of the given star (s=+1 in the direction of spin of the LMC disk), $i$ is the inclination of the LMC disk to the plane of the sky, $\rho$ is the radius coordinate expressed as angle on the sky, $\Phi$ is the position angle measured east of north, and $\Theta$ is the position angle of the line of nodes.  To determine $s$, we compute the angle of the proper motion vector $\Theta_t$ and compare it to the position angle $\Phi$, and set $s=-1$ if $90<(\Phi-\Theta_t)<270$ and $s=+1$ otherwise. 

The resulting deprojection shows \textit{roughly} the ordered rotational velocity in the inner $\sim10^{\circ}$ of the LMC. We refer to \citep[submitted]{Choi2022} for the detailed discussion about the stellar kinematics in the inner disk of the LMC. Beyond 8--10$^{\circ}$ from LMC center, on the other hand, the stars show a remarkable spread in $V_{int}$ values, as shown by the color scale markings of stars in \Cref{fig:lmc_vint}, with many of the various substructures discussed earlier showing markedly distinct, and even extreme, $V_{int}$ values. Meanwhile, stars in the region associated with ``Substructure 1'' have $V_{int}$ somewhat elevated above that for stars in the outer disk, whereas stars in the region associated with ``Substructure 2'' have very low, even negative $V_{int}$ relative to the outer disk.  

To demonstrate the dramatic change in the kinematical character of stars just beyond a radius of $\sim$10$^{\circ}$ in the southern LMC periphery, \Cref{fig:lmc_vint_dist} compares the distribution of $V_{int}$ values for stars within $8^{\circ}$ of the LMC center to those within the ``southeast periphery sector (SPS)''  outlined in \Cref{fig:lmc_vint} and spanning radii of 10-20$^{\circ}$. As stated before, stars within the SMC Exclusion Zone (red dotted line) are not included in the SPS. \Cref{fig:lmc_vint_dist} also includes as a control sample those stars in a similar range of radius but spanning the entire northern LMC periphery (the ``northern periphery sector (NPS)'').

As may be seen in \Cref{fig:lmc_vint_dist}, the SPS stars span a vastly broader range ($\sim$600 km s$^{-1}$) in $V_{int}$ than either the stars in the inner, disk-dominated region or in the NPS, which looks very much like the inner disk in terms of $V_{int}$ distribution. While some SPS stars share the nominal $V_{int}$ velocities of disk stars, the former are generally confined to SPS stars at smaller radius, as is evident in \Cref{fig:lmc_vint}.  On the other hand, a larger fraction of SPS stars have velocities with {\it more extreme} $V_{int}$ --- either much higher than the nominal LMC disk, or retrograde.  Neither of these types of $V_{int}$ are what is expected for the outermost parts of disks, where galaxy mass is typically distributed so that rotational velocity decreases with radius (but remains prograde).
Moreover, given that $V_{int}$ represents a 2-D, deprojected velocity to the LMC disk plane, not only does \Cref{fig:lmc_vint_dist} demonstrate just how ``non-disk-like'' are the motions of a large fraction of SPS stars, but it suggests that the full 3-D motions of some SPS may be even more distinct and extreme. That assessment is borne out by the stars in hand for which full 3-D motions are possible due to the availability of APOGEE radial velocities (RVs).

In Section 6 of \citet{Gaia2021}, a similar kinematical study of LMC outskirts is also reported. The authors pointed out that both the northern (northern tidal arm, NTA, in their paper) and southern substructure (southern tidal arm, STA, in their paper) have consistent velocities to those of LMC, and an additional substructure is detected to the east of LMC (ESS in their paper). While we agree that the northern substructure has consistent velocity distribution with those of the outer LMC disk, the southern substructure shows significant differences in velocity, especially an increase in stars with high in-plane velocities that is not present in the northern periphery region, and only a slightly larger velocity is detected to the east of LMC, which could be interpreted as an extension of the southern substructures.

\subsection{3D Motions and Metallicities for APOGEE Stars}

\begin{figure}
    \centering
    \includegraphics[width=\columnwidth]{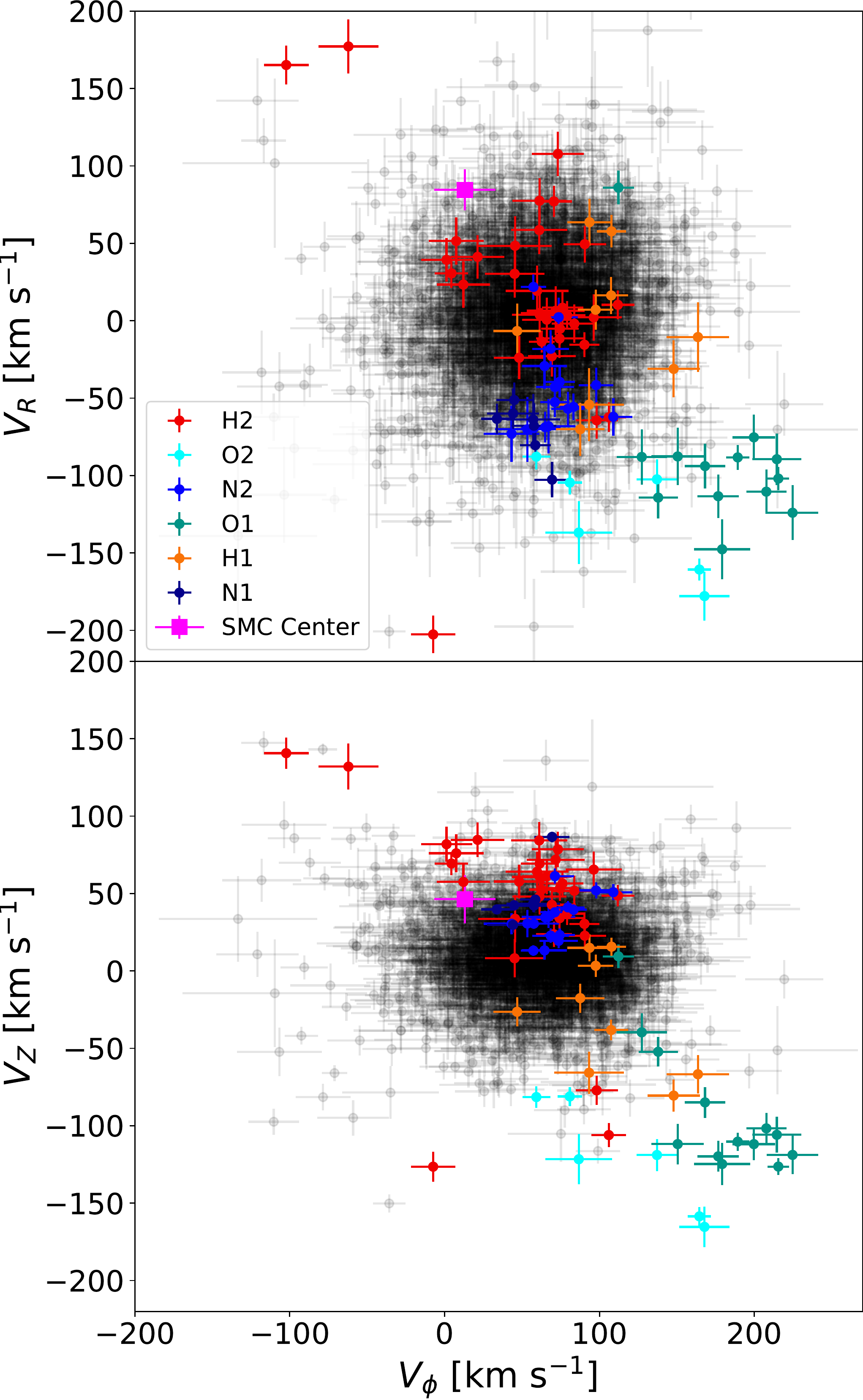}
    \caption{The vertical velocity ($V_Z$) and radial velocity ($V_R$) versus azimuthal velocity ($V_\phi$) of APOGEE DR17 stars,  with symbols colored for stars in the six outer LMC fields in the same way as in \Cref{fig:lmc_dist} and \Cref{fig:lmc_vint} and with LMC disk stars colored in grey as well as the SMC center as a magenta square. The error bars on individual velocities reflect only the measurement uncertainties in line-of-sight velocity and 2D proper motion.
    }\label{fig:lmcperi_3d_velocity}
\end{figure}

By combining APOGEE DR17 RVs with {\it Gaia} proper motions, full three-dimensional (3-D) motions can be calculated.  We use the same orientation of the LMC disk (i.e., line-of-nodes and inclination angle) as used in the model to calculate $V_{int}$ in Section \ref{sec:2Dmotion} to transform those 3-D motions into a cylindrical coordinate system appropriate to the LMC disk reference frame, where $V_R$ and $V_{\phi}$ are the radial and rotational motions projected onto the LMC disk plane and $V_Z$ is the motion perpendicular to the disk plane (where a positive $V_Z$ is towards the Sun). To perform this transformation, we first inverted Equation (5) from \citet{vanderMarel2002} to solve for $v_x^\prime$, $v_y^\prime$, and $v_z^\prime$ in the plane of LMC disk, computed the in-plane positions $x^\prime$ and $y^\prime$ using Equation (7) from \citet{vdM+Cioni2001}, and then computed $V_R$, $V_{\phi}$, and $V_Z$ as:

\[
V_R = (x^\prime v_x^\prime + y^\prime v_y^\prime)/R,
\]
\[
V_\phi = (y^\prime v_x^\prime - x^\prime v_y^\prime)/R,
\]
\[
V_Z = v_z^\prime.
\]

\Cref{fig:lmcperi_3d_velocity} shows the velocity distributions in this parameter space for each of the six individual APOGEE fields shown in \Cref{fig:lmc_dist}, along with stars from the LMC disk. This LMC disk sample is the same as in \citet{Nidever2020}. The latter stars define clear concentrations in velocity space.  It is immediately obvious that the stars in the O1 and O2 APOGEE fields have velocities very different from those of LMC disk stars, with strong (by more than 100 km s$^{-1}$), ``infalling'' radial motion and typically a faster $V_{\phi}$ (i.e., azimuthal) motion than that of LMC disk stars.  In the case of the O1 field the $V_{\phi}$ motions of some of the stars exceed that of the most rapidly rotating LMC disk stars by of order 100 km s$^{-1}$.  Given these quite different and extreme kinematics, it is difficult to conclude that the stars in the O1 and O2 fields are simple extensions of the LMC disk.  

\begin{figure}
    \centering
    \includegraphics[width=\columnwidth]{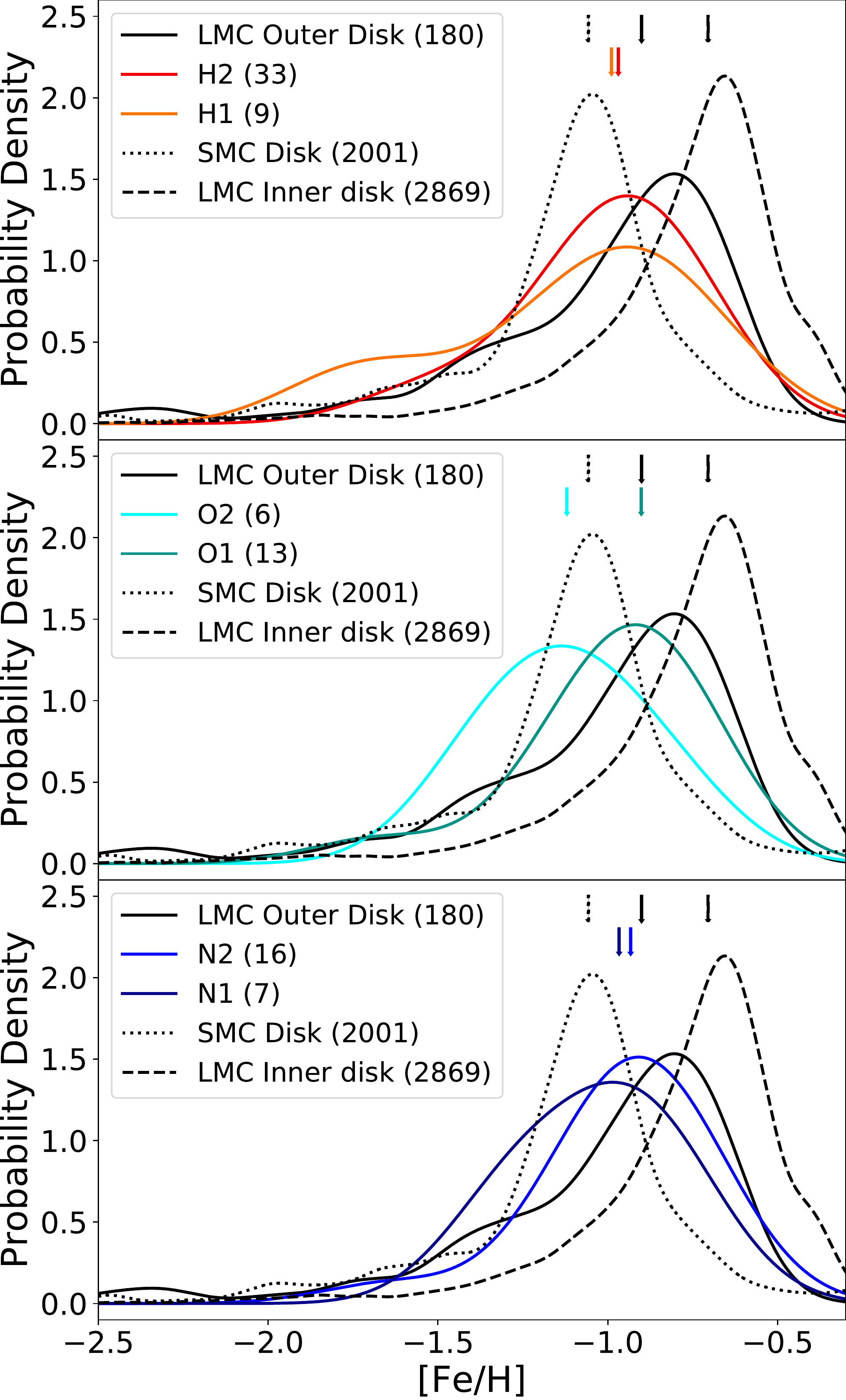}
    \caption{Probability density functions derived from KDE applied to the observed metallicities for the LMC inner disk (within 8 deg of the LMC center, black dashed lines), LMC outer disk (further than 8 deg from LMC center, black solid lines), SMC disk (within 7 deg from SMC center), and stars in all six APOGEE fields (same color with \Cref{fig:lmc_vint}). The median metallicity of each sample is labelled by the arrow close to the top of the figure. The number of stars within each sample is labelled in the legend of each panel.}\label{fig:lmcperi_feh}
\end{figure}

This assessment would seem to be at least partly supported by the spectroscopic metallicities for the stars in the O1 and O2 fields as derived by APOGEE. These are shown by the probability distribution functions derived by kernel density estimation (KDE) in  \Cref{fig:lmcperi_feh}.  As may be seen, the peak of the metallicity distribution function (MDF) for the O2 field is shifted by about 0.6 dex in [Fe/H] from that of the LMC inner disk. Even accounting for the gentle radial metallicity gradient in the LMC disk, the O2 field is still shifted by about 0.2 dex from the MDF of the outermost part of the disk (stars 8-10$^{\circ}$ from the LMC center, shown as the solid line in \Cref{fig:lmcperi_feh}). Indeed, the MDF for the O2 field is similar to, though slightly more metal poor than that of the SMC; however, while the O2 field lies very close to the SMC in the sky, the kinematics of the O2 and SMC stars are so disparate (e.g., separated by some 300 km s$^{-1}$ in the simple $V_{int}$ projection of proper motion; \Cref{fig:lmc_vint}) that it would seem to preclude a simple connection of the O2 stars to the SMC.

On the other hand, while the MDF of the O1 stars seems to match well that of the LMC outer disk, their 3-D motions are clearly quite distinct (\Cref{fig:lmcperi_3d_velocity}). However, all of these MDF comparisons must be considered tentative, given that there are only 13 and 6 stars with APOGEE data in each of the O1 and O2 fields, respectively. The metallicity and detailed chemical abundances of the stars in these six APOGEE fields are explored further in a companion paper by Mu\~noz-Gonzalez et al. (in preparation).

In contrast to the situation for the O1 and O2 fields, the stars in fields N1, N2, H1 and H2 do lie within the approximate 3-D velocity envelope of the LMC disk stars (\Cref{fig:lmcperi_3d_velocity}), albeit generally near the ``edge'' of the envelope.  This suggests a closer  connection of the stars in these APOGEE fields to the LMC disk.  This association is apparently supported by the MDFs of the various populations, in particular for the N1, N2 and H2 fields, which match well to the MDF of the outer disk.  Only the H1 MDF seems less consistent with the others, but this MDF consists of data for only seven stars. 
It is perhaps not so surprising that the N1 and N2 groups might be associated with the LMC disk, given that they lie right on the apparent spiral arm feature. However, these new APOGEE results suggest a closer connection of the two southern ``hook'' features to the LMC disk than previously thought. It also points to these two features as being exceptions to the bulk of the stars in the SPS region, which, based on their $V_{int}$ values, seem kinematically distinct from the LMC disk (Section \ref{sec:2Dmotion}).

\subsection{Comparison with Simulations}
To obtain better insight on the observed extreme in-plane velocities in the SPS region, we investigate the two simulations from \citet{Besla2012} of an interacting pair of LMC and SMC analogs, subject to the MW's gravitational potential under a first infall scenario. In these simulations, the LMC/SMC binary interaction produces tidal features qualitatively similar to what is broadly observed in the Magellanic system, and so are potentially useful for understanding our results on kinematic outliers. There is as of yet no consensus in the field regarding the recent interaction history between the Clouds \citep[e.g.,][]{Cullinane2021}. The major difference between the two \citet{Besla2012} simulations is the impact parameter of the most recent encounter ($\sim$100~Myr ago) between the Clouds: Model 1 has an impact parameter of $\sim$20~kpc, with consequently less dramatic effect on the structure of the galaxies, while Model 2 has an impact parameter of $\sim$2~kpc, with substantially more tidal debris at large distance from the parent bodies. 

To make our comparisons, we translate the 6D phase space information of the simulated LMC/SMC stellar particles to the observed frame. More specifically, we recenter all simulated LMC/SMC particles to match the observed center of mass position and velocity vectors of the LMC, ($X,Y,Z$) = (-1, -41, -28)~kpc and ($V_X$,$V_Y$,$V_Z$) = (-57, -226, 221)~km~s$^{-1}$ \citep{Kallivayalil2013}. This step is necessary as these simulations were designed such that 3D velocity vector of the LMC matched that measured earlier by \citet{Kallivayalil2006}. This shift is applied to the entire simulated Magellanic system and does not change any of the motions of stellar particles internal to each simulated galaxy. We then translate the positions and velocities of each star particle from the Galactocentric coordinate system to $\alpha$, $\delta$, line-of-sight distance, \pmra, \pmdec, and line-of-sight velocities using the Python library \texttt{astropy.coordinates}. We note that we exclude any star particles younger than 1 Gyr old from our analysis, in order to enable comparisons to the observational results based on RGB stars.

We apply the same kinematic modeling procedure to these simulated LMC disk star particles as we did for the data (Section \ref{sec:results}), which result in fitted parameters for both the bulk center-of-mass motion and internal rotation of the LMC in the two simulations. We then apply these model parameters to all star particles in the simulations, including the SMC particles, returning values for $V_{int}$, $V_R$, $V_{\phi}$, and $V_Z$ for all particles with respect to the LMC center-of-mass reference frame. 

We note that the inclination and line-of-node position angle of the simulated LMC disk in Model 1 and Model 2 are not an exact match to the observed values with regards to our line of sight \citep[see Section 3.2 in][]{Besla2016}, and that the center of mass position and velocity of the simulated SMC is not exactly matched to the observed values as described in \citet{Besla2012}. However, no corrections are made to the simulated LMC and SMC to make them consistent with these two observed values, as the velocities that we care about are all relative to the LMC center-of-mass reference frame. Thus, the analysis of simulations presented here is only to serve as a proof of concept for the plausible range of kinematics associated with stellar debris tidally removed from the LMC-SMC interactions.

\begin{figure*}
    \centering
    \includegraphics[width=\textwidth, trim=3cm 1cm 3cm 1cm, clip=True]{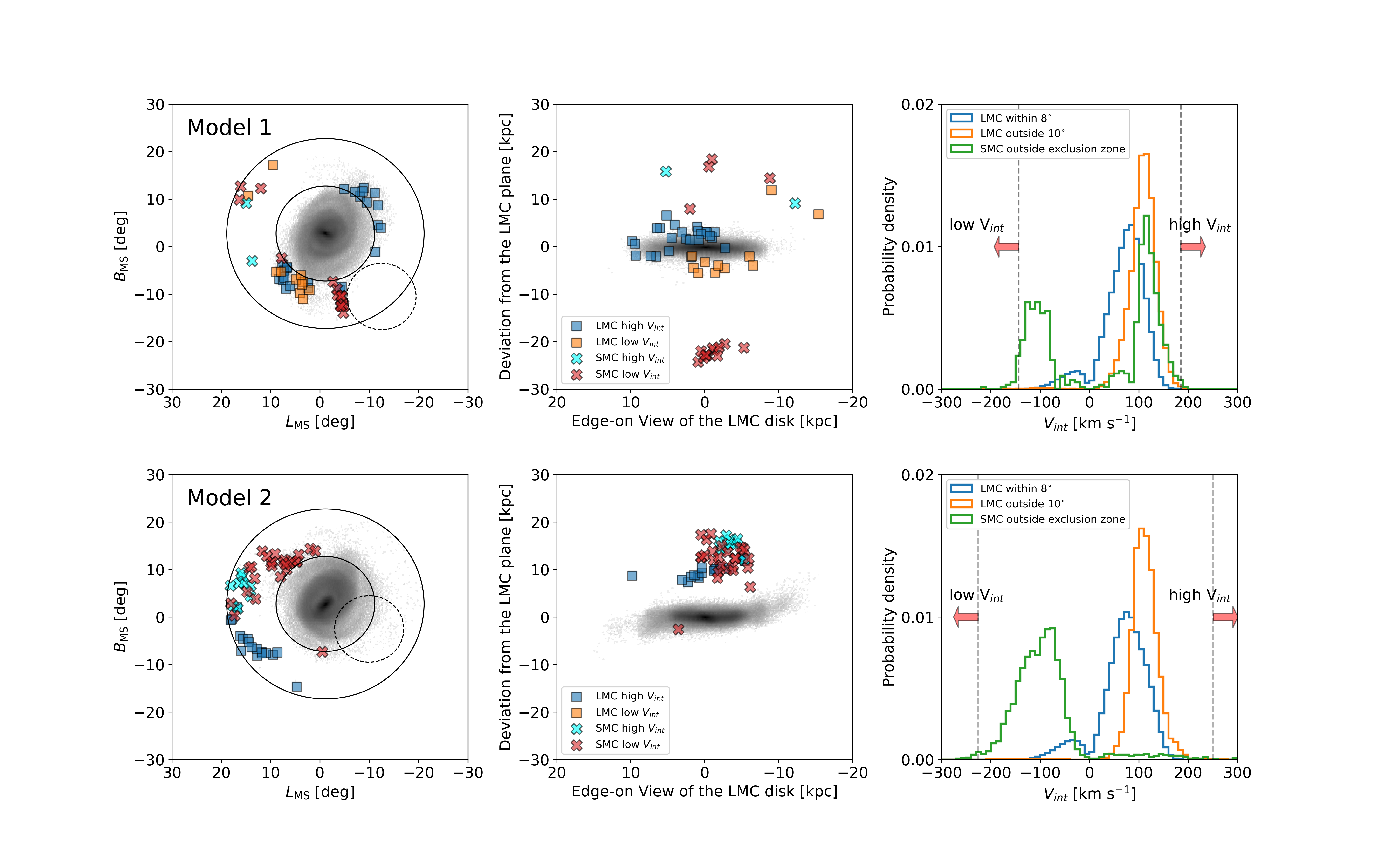}
    \caption{\textit{Top row (Model 1):} the 2D star count map of the simulated LMC (left column) and the edge-on view of the simulated LMC (middle column) overplotted with the kinematic outliers with high and low $V_{int}$. As shown in the right column, the kinematic outliers are identified as 0.15\% population in the low and high tails of the $V_{int}$ distribution of all the star particles that are within the 10-20$^{\circ}$ annulus, but outside the SMC exclusion zone. These outliers preferentially reside in tidally-induced low-density structures and are found above and below the main disk plane. While high $V_{int}$ star particles mainly originate in the LMC, low $V_{int}$ star particles have both LMC and SMC origin. \textit{Bottom row (Model 2):} The panels are the same as for Model 1. Similar to Model 1, the kinematic outliers are found in low surface brightness tidal features, but with a more skewed spatial distribution. Model 2 shows much stronger extraplanar features; most outliers reside 10-20~kpc above the main disk. }\label{fig:lmc_model_vint}
\end{figure*}

\Cref{fig:lmc_model_vint} presents the kinematic properties of Model 1 (upper panels) and Model 2 (lower panels). We apply the same spatial cuts as described in Section~\ref{sec:data}. Specifically, we define the LMC main disk as the inner 8$^{\circ}$ from the LMC center, focus on the 10-20$^{\circ}$ annulus to look for kinematically distinct populations, and exclude the SMC particles within 7$^{\circ}$ from the SMC center in our analysis. We also exclude those SMC particles that are outside the SMC exclusion zone but within the 10-20$^{\circ}$ annulus if they have proper motions inconsistent with the majority of the LMC particles. Due to the inconsistent line-of-node position angles of the simulated LMC disks with that of the observed disk, we analyze the 10-20$^{\circ}$ annulus as a whole instead of dividing the annulus into two sectors as we did for the observation (North vs. South sectors). 

From the $V_{int}$ distribution of all the LMC/SMC star particles within the 10-20$^{\circ}$ annulus (except for the SMC particles inside the SMC exclusion zone), we identify kinematic outlier stars as those that have $V_{int}$ below the 0.15 percentile value (low $V_{int}$ stars) or above the 99.85 percentile value (high $V_{int}$ stars). This is equivalent to 3-sigma outlier selection for the case of a normal distribution. We mark the low/high $V_{int}$ values for Model 1 (-143/185~km~s$^{-1}$) and Model 2 (-226/251~km~s$^{-1}$) in the upper and lower right panels, respectively. The mass fraction of kinematic outlier stars in the 10-20$^{\circ}$ annulus relative to the stars in the LMC main disk is $\sim$0.0043\% in both Model 1 and 2. If we do the same outlier selection for the observational data, the computed low and high $V_{int}$ values are -254~km~s$^{-1}$ and 319~km~s$^{-1}$, respectively. Model 2, which has a closer LMC/SMC impact parameter than Model 1, shows a better agreement with the observation in terms of the low and high $V_{int}$ values. However, even Model 2 cannot reach $V_{int}$ values as high as those observed, indicating that a stronger tidal perturbation might be needed to reproduce the extreme velocity stars seen in the observation. The number fraction (which is a proxy for mass fraction by virtue of the fact that RGB stars have similar masses) of the kinematic outliers among the Gaia-selected RGB stars relative to the those within the inner 8$^{\circ}$ is $\sim$0.0045\%, which might be considered a rough upper limit because the RGB selection is likely not 100\% complete in the innermost region due to crowding \citep{Gaia2021}. However, it is notable that the simulations contain roughly the same fraction of kinematic outliers as the observations.  

The upper and lower left panels in \Cref{fig:lmc_model_vint} show the spatial distribution of kinematic outlier star particles on the 2D star count maps of the simulated LMC from Model 1 and Model 2, respectively. The two solid black circles denote the radii of 10$^{\circ}$ and 20$^{\circ}$ from the LMC center, while the blue dashed circle marks the radius of 8$^{\circ}$ from the LMC center. The green dashed line shows the SMC exclusion zone. The population consisting of the kinematic outliers in the 10-20$^{\circ}$ annulus for each model is different. In Model 1, most of the high  $V_{int}$ stars in the annulus have an LMC origin, while the low $V_{int}$ stars have both LMC and SMC origin. In Model 2, all the low $V_{int}$ stars in the annulus are SMC debris. In general, all the kinematic outliers in both Model 1 and 2 are found around tidally induced low surface brightness features. However, the detailed spatial distributions of kinematic outliers are different in the two models. Model 1 shows a rough bi-polar distribution, while Model 2 shows a one-sided distribution. In 3D space, the kinematic outliers are mostly extraplanar, as is clearly seen in the edge-on view of the simulated LMC disks (middle panels). In Model 1, the majority of the outliers with LMC origin are found both above and below the main disk, but within $\sim$10~kpc. On the other hand, the outliers with SMC origin are located far above or below the main LMC disk. In Model 2, almost all of the kinematic outliers, including the LMC debris, are $\sim$10-20~kpc above the LMC main disk. 

On the recommendation of the anonymous referee, we also examined plots of component velocities $V_Z$ and $V_R$ versus $Z$ for the simulations, and compared features found in them to those selected by $V_{int}$.  We find that our $V_{int}$ selection identifies features that would also be seen as outliers in these plots of component velocity versus $Z$; the advantage of $V_{int}$ is that we can compute its value for the observations, whereas we have no way to measure $Z$, and thus must assume that $Z=0$ for all stars.

\begin{figure}
    \centering
    \includegraphics[width=\columnwidth, trim=2cm 2cm 2cm 1cm, clip=True]{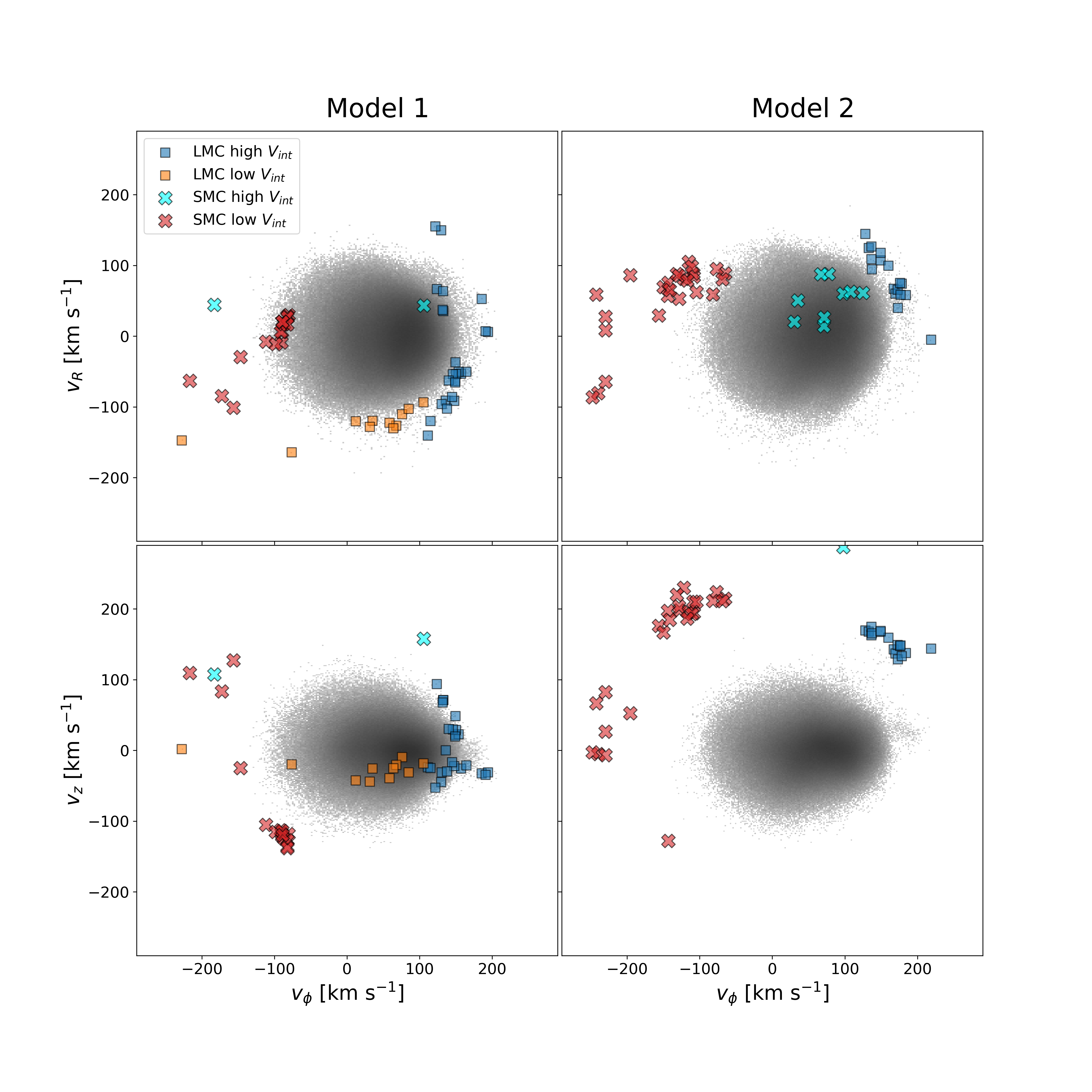}
\caption{The vertical velocity ($V_Z$) and radial velocity ($V_R$) versus azimuthal velocity ($V_\phi$) distributions for the simulated LMC from Model 1 (left columns) and Model 2 (right columns). Kinematical outliers from the LMC (blue and red squares) and SMC (cyan and orange crosses) are highlighted. The coloring scheme for kinematical outliers is the same as in \Cref{fig:lmc_model_vint}. As seen for the stars in the APOGEE O1 and O2 fields (see \Cref{fig:lmcperi_3d_velocity}), most of the $V_{int}$-selected kinematical outliers also have extreme individual velocity components compared to the stars in the main disk. }\label{fig:lmc_model_3comp}
\end{figure}

In \Cref{fig:lmc_model_3comp}, we show the $V_R$, $V_{\phi}$, and $V_Z$ velocity components for the two models. The underlying gray scale shows the velocity distributions of star particles in the LMC disk within 8$^{\circ}$. We overplot the kinematical outliers shown in \Cref{fig:lmc_model_vint} using the same color and symbol schemes. Similar to what we see from the stars in the O1 and O2 APOGEE fields (\Cref{fig:lmcperi_3d_velocity}), the kinematical outliers in the simulations show distinct behaviors from the majority of the star particles in the main disk. It is difficult to make a fair comparison between the observations and the simulations because the O1 and O2 APOGEE fields probe a tiny portion of the 10-20$^{\circ}$ annulus with a narrow coverage of position angles ($\sim$10$^{\circ}$ around the position angle of 180$^{\circ}$), whereas the kinematical outliers in the simulations are tied to a larger range of position angles. Nevertheless, the amplitudes of offsets in each velocity component from the majority of the LMC disk star particles in the models are comparable to those seen in the observations. 

\section{Discussion and Conclusion} \label{sec:conclusion}

From our analysis of the 2-D velocities based on {\it Gaia} proper motions for a large number of stars --- from which we derive a deprojected, in-plane velocity, $ V_{int}$, per star --- combined with 3-D velocities and metallicities for a smaller collection of stars in new APOGEE fields, we find and conclude the following:

\begin{itemize}
    \item The periphery of the LMC contains stars from a variety of origins and with a clear north-south dichotomy:  The stars in the northern LMC periphery (represented by stars in the NPS region of proper motions and the N1 and N2 APOGEE fields) seem to have ties to the outer LMC disk, based on both their kinematics and MDFs.  In contrast, the stars in the southern LMC periphery (represented by those in the SPS region generally) show a more heterogeneous MDF and an especially diverse kinematical character, with the latter exhibiting a remarkably extreme range in velocities, with some stars sharing the motions of the LMC disk, but a significant fraction of stars moving quite unlike the stars in the LMC disk.  
    \item Within the SPS region, the areas represented by the hook-like features previously identified by \citet{Mackey2018} have $V_{int}$ values more like those found in the LMC outer disk, and this kinematical association is supported by the observations of stars in the H1 and H2 fields, which show 3-D velocities and MDFs like those of the outer LMC disk.
    \item On the other hand, stars at larger radius in the SPS contain stars with more extreme kinematics (showing both retrograde velocities and prograde velocities at much higher velocity than the LMC disk), as exemplified by the 3D motions of the stars in the O1 and O2 fields, which cannot be viewed as a simple dynamical extension of the LMC disk.
    \item The stars in the O2 field have a spatial and metallicity distribution suggesting a connection to the SMC, but a velocity character extremely distinct from the SMC.  Meanwhile, stars in the O1 field have an MDF resembling that of the outer LMC disk, but, again, a kinematical character quite distinct from that association.  For these stars, one possibility is that they are highly disturbed tidal debris from the LMC/SMC interaction, which we explore by comparing their kinematical nature with those from hydrodynamical N-body simulations (see below).  However, we cannot rule out that some APOGEE stars in these fields are of an ``external'' origin, the LMC-equivalent of accreted halo substructure, evidence for which has previously been suggested by \citet{Majewski2009}.
\end{itemize}

From our comparisons with two hydrodynamical N-body simulations of an interacting LMC-SMC system \citet{Besla2012}, we find and conclude the following:
\begin{itemize}
    \item The observed extreme velocity stars can be qualitatively reproduced by the tidal interactions between the LMC and SMC. The kinematical outliers identified in the simulations are extraplanar and preferentially found in tidally-induced low density features. This suggests that many of the stars in the SPS region are also out of the plane of the LMC.
    \item The detailed populations of the kinematical outliers depend on the interaction histories. In Model 1, where there is no direct collision between the MCs, the contribution of the SMC particles to the high positive in-plane velocity population is negligible. In Model 2, where a recent direct collision occurred between the MCs, there is no contribution of the LMC particles to the high negative in-plane velocity population. We note that the LMC (SMC) debris are dominant components of the high positive (negative) in-plane velocity population in both models.
    \item Although the simulations are able to provide a plausible explanation for the kinematical properties of extreme velocity stars, neither models reproduce the details of the observed $V_{int}$ distribution, including the extended high positive in-plane velocity tail seen in the observation. This might suggest that future models need a stronger perturbation (e.g., heavier SMC) to reach the observed highest $V_{int}$ values. To test this, exploring a much broader parameter space for the interaction history is required.
\end{itemize}

Obviously, additional investigation is needed to solidify these conclusions.  Larger spectroscopic samples would, of course, be a great help.  But other data exist now that might help with firming up or ruling out the above conclusions.  One particularly useful aid would be the discernment of relative distances of the LMC disk, SMC disk, and the periphery field stars, which, combined with the relative motions, would provide more definitive conclusions regarding the origin of the various spatio-kinematically distinct features. Unfortunately, at present the uncertainties associated with distance gauging individual sources at these great separations from us are still too large.  We attempted to statistically assess the relative distances of stars based on color-magnitude distributions, but confess that these investigations proved quite inconclusive. 

Additional evidence bearing on possible associations of Magellanic periphery stars with either the LMC or SMC would come from comparisons of detailed chemical abundance patterns, which, conveniently, are provided by the APOGEE database.  In a companion paper (Mu\~noz et al. 2021) we undertake just such an analysis.

\acknowledgments

X.C., S.R.M. and B.A. acknowledge support from National Science Foundation (NSF) grant AST-1909497, while D.L.N. acknowledges NSF grant AST-1908331. A.M. acknowledges support from FONDECYT Regular grant 1212046 and funding from the Max Planck Society through a “PartnerGroup” grant. C.M. thanks the support provided by FONDECYT Postdoctorado No.3210144.

This work has made use of data from the European Space Agency (ESA) mission {\it Gaia} (\url{https://www.cosmos.esa.int/gaia}), processed by the {\it Gaia} Data Processing and Analysis Consortium (DPAC, \url{https://www.cosmos.esa.int/web/gaia/dpac/consortium}). Funding for the DPAC has been provided by national institutions, in particular the institutions participating in the {\it Gaia} Multilateral Agreement.

Funding for the Sloan Digital Sky Survey IV has been provided by the Alfred P. Sloan Foundation, the U.S. Department of Energy Office of Science, and the Participating Institutions. 

SDSS-IV acknowledges support and resources from the Center for High Performance Computing  at the University of Utah. The SDSS website is www.sdss.org.

SDSS-IV is managed by the Astrophysical Research Consortium for the Participating Institutions of the SDSS Collaboration including the Brazilian Participation Group, the Carnegie Institution for Science, Carnegie Mellon University, Center for Astrophysics | Harvard \& Smithsonian, the Chilean Participation Group, the French Participation Group, Instituto de Astrof\'isica de Canarias, The Johns Hopkins University, Kavli Institute for the Physics and Mathematics of the Universe (IPMU) / University of Tokyo, the Korean Participation Group, Lawrence Berkeley National Laboratory, Leibniz Institut f\"ur Astrophysik Potsdam (AIP),  Max-Planck-Institut f\"ur Astronomie (MPIA Heidelberg), Max-Planck-Institut f\"ur Astrophysik (MPA Garching), Max-Planck-Institut f\"ur Extraterrestrische Physik (MPE), National Astronomical Observatories of China, New Mexico State University, New York University, University of Notre Dame, Observat\'ario Nacional / MCTI, The Ohio State University, Pennsylvania State University, Shanghai Astronomical Observatory, United Kingdom Participation Group, Universidad Nacional Aut\'onoma de M\'exico, University of Arizona, University of Colorado Boulder, University of Oxford, University of Portsmouth, University of Utah, University of Virginia, University of Washington, University of Wisconsin, Vanderbilt University, and Yale University.

\vspace{5mm}
\facilities{}

\software{astropy \citep{astropy}, topcat \citep{topcat}}

\clearpage
\bibliography{sample63}{}
\bibliographystyle{aasjournal}

\end{document}